\documentclass[prb,twocolumn,aps,showpacs,superscriptaddress]{revtex4}
\usepackage{graphicx}
\def\ltsim{\vbox {\hbox{\lower .8\baselineskip \hbox{$<$}} \break
                 \hbox{\lower 0.2\baselineskip \hbox{$\sim$}} } }

\begin{document}

\title{A mesoscopic Resonating Valence Bond system on a triple dot}

\author{Karyn Le Hur}
\affiliation{D\' epartement de Physique et RQMP, Universit\'e de Sherbrooke, Sherbrooke, Qu\' ebec, Canada, J1K 2R1}
\author{Patrik Recher}
\altaffiliation[Also at ]{Institute of Industrial Science,
University of Tokyo, 4-6-1 Komaba, Meguro-ku, Tokyo 153-8505,
Japan} \affiliation{Quantum Entanglement Project, E.L. Ginzton
Laboratory, SORST, JST, \\Stanford University, Stanford,
California 94305-4085, USA}
\author{\' Emilie Dupont}
\affiliation{D\' epartement de Physique et RQMP, Universit\'e de
Sherbrooke, Sherbrooke, Qu\' ebec, Canada, J1K 2R1}
\author{Daniel Loss}
\affiliation{Department of Physics and Astronomy, University of
Basel, Klingelbergstrasse 82, CH-4056 Basel, Switzerland}

\begin{abstract}
We theoretically introduce a mesoscopic pendulum from a triple dot. The pendulum
is fastened through a singly-occupied dot (spin qubit). Two other
strongly capacitively coupled islands form a double-dot charge
qubit with one electron in excess oscillating between the two
low-energy charge states $(1,0)$ and $(0,1)$. The triple dot is placed between two
superconducting leads. Under realistic conditions, the main
proximity effect stems from the injection of resonating singlet
(valence) bonds on the triple dot. This gives rise to a Josephson
current that is charge- and spin-dependent and, as a consequence,
exhibits a distinct resonance as a function of the superconducting
phase difference.
\end{abstract}
 \date{\today}
\maketitle

By analogy with quantum optics, the production of entangled states
in condensed matter devices have inherently emerged as a
mainstream in nanoelectronics\cite{Nielson}. Entanglement between
electrons, besides checking fundamental quantum properties such as
non-locality, could be exploited for building logical gates and
quantum communication devices.  The realization of electron
entangled states might result from strong interactions in
nanoscopic systems. Mostly, this offers a room to treat the spin
in a quantum dot as a qubit\cite{Loss} with generally a quite long
decoherence time\cite{Kouw,Marcus}. Spin entanglement scenarios
have been envisioned in such a framework\cite{Ent}. Theoretical research on
the possibility to control and detect the spin of electrons
through their charges has also blossomed
recently\cite{Lesovik,Buttiker,Saraga}. Let us recall that the
direct coupling of two quantum dots by a tunnel junction might be
used to create entanglement between spins\cite{Loss} and such spin
correlations might be observed in transport experiments\cite{Loss1}.
Another mechanism to induce spin correlations would consist
to place a double quantum dot away from resonance in a vertical
configuration between two superconducting (SC) leads\cite{Choi}.
We go beyond and (theoretically) explore other entanglement mechanisms based on
triple dot devices and more precisely on the prolific proximity
between a {\it spin} and a double-dot {\it charge} qubit. 
We consider the idea to realize singlets resonating
between equivalent low-energy configurations on a triple dot by
placing the latter between two SC electrodes. This can be viewed
as a mesoscopic Resonating Valence Bond (RVB)
system\cite{Anderson} resulting in a Josephson current with
a distinct resonant-like profile as a function of the superconducting phase
difference.

The pillar of this mesoscopic RVB system is what we refer to as the mesoscopic pendulum. We take two strongly capacitively coupled quantum dots (say, dots 1 and 3 of Fig. 1) that form a charge qubit; only one electron in excess is permitted on dots 1 and 3 which thus embodies the weight of the pendulum. The pendulum is fixed through the dot 2 which is singly-occupied and off resonance (spin qubit).
\begin{figure}[ht]
\begin{center}
\includegraphics[width=6.4cm,height=4.8cm]{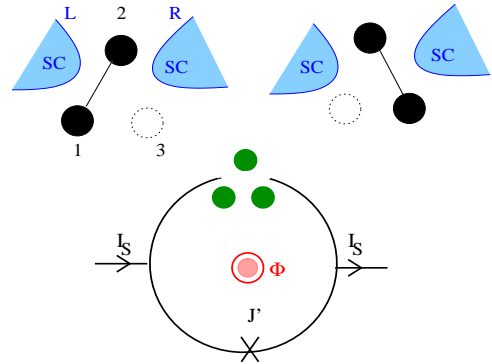}
\end{center}
\vskip -0.7cm
\caption{(Color online) The pendulum with resonating singlet bonds under consideration: The dots 1 and 3 are strongly capacitively coupled and form a two-level system characterized by the degenerate orbital states $(1,0)$ or $(0,1)$. Dot 2 is singly occupied and the direct tunnel coupling between dots 1 and 2 (or 2 and 3) is negligible. The spin entanglement induced by the SC leads between dot 2 and, say, dot 1 (or dot 3), embodies the rod of the pendulum.  The singlets can also resonate through the direct tunnel coupling ${\cal T}'$ between dots 1 and 3. Consequences in a SQUID geometry are properly analyzed.}
\end{figure}
Then, we place the quantum pendulum between two SC leads. The rod of the pendulum in Fig. 1 represents the emergence of spin entanglement between dot 2 and, say, dot 1, induced by the proximity from the SC leads. Through the tunnel coupling ${\cal T}'$ between dots 1 and 3 and the closeness of the SC leads, hence the singlet bonds will resonate on the triple dot.

{\it Model}.--- Assuming that the capacitive coupling between dot
1 (dot 3) and dot 2 is negligible (that is quite legitimate for
the situation of Fig. 1), the general charging energy for the dots
1 and 3 takes the standard form\cite{Wiel} $E_{n_1
n_3}=E_{c1}(n_{g1}-n_1)^2
+E_{c3}(n_{g3}-n_3)^2+E_{c13}(n_{g1}-n_1)(n_{g3}-n_3)$, where we
have introduced the charging energies $E_{c1,3}=e^2C_{\Sigma
3,1}/[2(C_{\Sigma 1} C_{\Sigma 3}-C_{13}^2)]$,  $E_{c13}=e^2
C_{13}/(C_{\Sigma 1} C_{\Sigma 3}-C_{13}^2)$. Here,
$n_{g1,3}=C_{g1,3} V_{g1,3}/e$ are the mean numbers of holes on
the gates --- coupled to dots 1 and 3 through the capacitances
$C_{g1,3}$ ---  being used to change the numbers of electrons
$n_1$ and $n_3$ on dots 1 and 3 through the gate voltages
$V_{g1,3}$, $C_{13}$ stands for the capacitive coupling between
dots 1 and 3, and $C_{\Sigma 1,3}=C_{g1,3}+C_{13}$. Of interest to
us is the strong inter-dot capacitive limit $C_{13}\sim C_{g1,3}$
with $n_{g1}\sim n_{g3}\sim 1/2$ such that $E_{01} \sim E_{10} \ll
(E_{00},E_{11})\ll (E_{20},E_{02})$. The low energy physics can
thus be studied within the restricted Hilbert space in which only
the $(n_1,n_3)=(0,1)$ and $(1,0)$ states are allowed for the dots
1 and 3. The manipulation of a single charge in a double dot is
now well accessible experimentally\cite{Hayashi,Dicarlo,Petta,Jeong}. A similar regime
can be reached with SC dots\cite{Pashkin,Levy}. We assume that
the dots 1 and 3 form a {\it nonlocal charge} qubit (with only one
electron in excess) and thus we can resort to the projecting
operators $\hat{P}_{01}$ and $\hat{P}_{10}$ to project on the
$(0,1)$ and $(1,0)$ states respectively\cite{Meirong}. Below, we
refer to ${\cal H}_{13}$ as the charging Hamiltonian for dots 1 and 3 related to $E_{n_1 n_3}$.

The dot 2 contains one electron in excess at the energy $\epsilon_2=-\epsilon$ with $\epsilon>0$ ({\it spin} qubit) and is subject to a strong on-site Coulomb repulsion $U$ which is typically the charging energy on dot 2, leading to the Hamiltonian
\begin{equation}
{\cal H}_{2} = \epsilon_2 \sum_{\sigma} d^{\dagger}_{2\sigma} d_{2\sigma} + Ud^{\dagger}_{2\uparrow} d_{2\uparrow}d^{\dagger}_{2\downarrow} d_{2\downarrow}.
\end{equation}
The SC leads are described by the BCS Hamiltonian
\begin{eqnarray}
{\cal H}_{SC} = \sum_{j=L,R} \int_{\Omega_j}\frac{d{\bf r}}{\Omega_j}
\sum_{\sigma=\uparrow,\downarrow} \Psi^{\dagger}_{\sigma}({\bf r}) x({\bf r}) \Psi_{\sigma}({\bf r}) \\
\nonumber
+\Delta_j({\bf r})\Psi^{\dagger}_{\uparrow}({\bf r}) \Psi^{\dagger}_{\downarrow}({\bf r}) +\hbox{H.c.},
\end{eqnarray}
where $\Omega_j$ is the volume of lead $j$, $x({\bf r})=(-i\hbar\nabla+\frac{e}{c}{\bf A})^2/(2m)-\mu$, and $\Delta_j({\bf r})=\Delta_j e^{-i\phi_j({\bf r})}$ is the pair potential. For simplicity, we assume identical leads with the same chemical potentials $\Delta\mu=0$ and $\Delta_L=\Delta_R=\Delta$.
Finally, the relevant tunneling Hamiltonians take the form
\begin{eqnarray}
{\cal H}_{\cal T} &=& \sum_{j=L,R;\sigma=\uparrow,\downarrow} {\cal T}_2 e^{i\phi({\bf r}_{j}-{\bf r}_2)} \Psi^{\dagger}_{\sigma}({\bf r}_j) d_{2\sigma} +\hbox{H.c.}
\\ \nonumber
&+&\sum_{\sigma=\uparrow,\downarrow}{\cal T}_1 e^{i\phi({\bf r}_{L}-{\bf r}_1)} \Psi^{\dagger}_{\sigma}({\bf r}_L) d_{\sigma}\hat{P}_{10}+\hbox{H.c.} \\ \nonumber
&+& \sum_{\sigma=\uparrow,\downarrow}{\cal T}_3 e^{i\phi({\bf r}_{R}-{\bf r}_3)} \Psi^{\dagger}_{\sigma}({\bf r}_R) d_{\sigma}\hat{P}_{01}+\hbox{H.c.}. \\ \nonumber
{\cal H}_{{\cal T}'} &=& |{\cal T}'|e^{i\lambda}e^{i\phi({\bf r}_{3}-{\bf r}_1)} \hat{Q}^+ + \hbox{H.c.}.
\end{eqnarray}
We have introduced the Aharonov-Bohm phase $\phi({\bf r}_n - {\bf
r}_m)= -\frac{\pi}{\Phi_0}\int_{{\bf r}_m}^{{\bf r}_n} d{\bf
l}\cdot{\bf A}$ and $\Phi_0=hc/2e$ is the SC flux quantum. Here,
the symbols ${\bf r}_1$, ${\bf r}_2$, and ${\bf r}_3$ refer to the
positions of the dots 1,2, and 3, respectively whereas ${\bf r}_L$
and ${\bf r}_R$ embody the coordinates in the SC leads L and R.
Note that in the present setup, an electron on dot 1 can either
tunnel into the SC lead $L$ or still onto dot $3$ via ${\cal
T}'=|{\cal T}'|\exp(i\lambda)$. Similarly, an electron on dot 3
can either tunnel into the SC lead $R$ or on dot 1. The tunnel
coupling between dot 2 and, say, dot 1 (or dot 3), is negligible.
Since there is a single electron in excess delocalized between
dots 1 and 3 we find it convenient to introduce a unique electron
annihilation operator $d_{\sigma}$ such that $\sum_{\sigma}
d^{\dagger}_{\sigma} d_{\sigma} =1$,
$d_{\sigma}\hat{P}_{10}=d_{1\sigma}\hat{P}_{10}$, and
$d_{\sigma}\hat{P}_{01}=d_{3\sigma}\hat{P}_{01}$, with
$d_{1\sigma}$ $(d_{3\sigma})$ annihilating explicitly an electron
on dot $1$ $(3)$. Along the lines of Refs.
\onlinecite{Meirong,Borda}, we have also exploited the double-dot
charge qubit notations for ${\cal H}_{{\cal T}'}$. The raising
operator $\hat{Q}^+$ --- acting exclusively on the state $|n_1
n_3\rangle$ --- ensures that each time an electron travels from
dot 1 to dot 3 this causes a flip $(1,0)\rightarrow (0,1)$ that
means $\hat{Q}^+|10\rangle = |01\rangle$,
$\hat{Q}^-|01\rangle=|10\rangle$,
$\hat{Q}_z=(\hat{P}_{01}-\hat{P}_{10})/2$, and $\hat{Q}^+
|01\rangle = 0$. Bear in mind that, for ${\cal H}_{\cal T}=0$, the
double-dot charge qubit is embodied by the ground-state wave
function\cite{Wiel}  $|\Psi_Q\rangle = \sqrt{P_{10}}|10\rangle
-\sqrt{{P}_{01}}e^{-i\phi({\bf r}_1-{\bf
r}_3)+i\lambda}|01\rangle$ with $P_{10}=\langle
\Psi_Q|\hat{P}_{10}|\Psi_Q \rangle = \langle \hat{P}_{10} \rangle$
and $P_{01}=\langle \Psi_Q|\hat{P}_{01}|\Psi_Q \rangle$ satisfying
$\langle \Psi_Q|{\cal H}_{\cal T'}|\Psi_Q\rangle =-|{\cal T'}|$
when $P_{01}=P_{10}=1/2$. Below, we consider ${\cal T}_1 = {\cal
T}_2 = {\cal T}_3 = {\cal T}$. The total Hamiltonian reads ${\cal
H}={\cal H}_0+{\cal H}_{\cal T}+{\cal H}_{{\cal T}'}$ with ${\cal
H}_0={\cal H}_{13}+{\cal H}_2+{\cal H}_{SC}$.

{\it Charge qubit at resonance}.--- Let us start with
$E_{01}=E_{10}=E_Q$. The low-energy subspace is made of one
localized electron on dot 2 and one electron delocalized on dots 1
and 3. Now we derive an effective Hamiltonian ${\cal H}^{eff}$
respecting this reduced Hilbert space. The low-energy states are
well separated by the superconducting gap $\Delta$ as well as the
Coulomb repulsions $U$  and $E_{20}$ or $E_{02}$. We assume that
dot 2 is small enough such as $U-\epsilon\gg \epsilon$ to hinder a
double occupancy on dot 2; by construction $(E_{20},E_{02})\gg
(E_{11},E_{00 })$. We introduce $P$ the projection operator on the
lowest-energy subspace and $K=(1-P)$ the projection operator on
excited states:
\begin{equation}
\label{EffH} {\cal H}^{eff} = P {\cal H} P +P{\cal
H}K\frac{1}{E-K{\cal H}K}K{\cal H}P.
\end{equation}
An expansion in ${\cal H}_{\cal T}$ can be performed and we identify $E$ with $E_Q-\epsilon$, {\it i.e.}, with the ground-state energy associated to ${\cal H}_0$. We envision the limit where $\epsilon/\Delta\ll 1$, nevertheless $\epsilon>0$ to ensure a trapped electron on dot 2,  and $E_{00}-E_Q\ll \Delta$ such that we can disregard individual qubit (and Kondo) contributions. This realm can be achieved
experimentally by adjusting the gate voltages of the dot 2 and of the double-dot charge qubit.
Temperatures are less than $\epsilon$ and $(E_{00}-E_{Q})$ (close to  zero).

\begin{figure}[ht]
\begin{center}
\includegraphics[width=4.2cm,height=4.2cm]{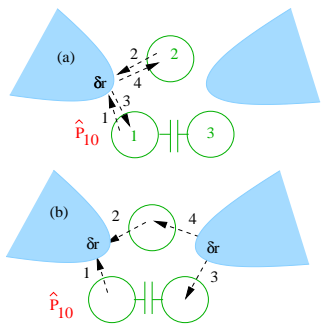}
\end{center}
\vskip -0.4cm \caption{(Color online) Main nonlocal Andreev
mechanisms when the electron on the double-dot charge qubit is
initially on dot 1 and $E_{10}\sim E_{01}=E_Q$. The numbered
arrows indicate the direction and the order of the electron
transfers. Since $\epsilon\ll \Delta$ and $E_{00}-E_Q\ll \Delta$,
we ignore {\it individual} contributions\cite{Choi} from dot 2
giving a $\pi$-junction\cite{Glazman} and from the charge
qubit\cite{Borda}.}
\end{figure}

The main proximity effects with the SC leads containing
$\hat{P}_{10}$ are the nonlocal Andreev tunneling processes of
Figs. (2a) and (2b) resulting in a nonlocal spin-entangled
electron state;  the second-order contribution gives a constant.
Assuming $E_{00}\ll E_{11}$ we omit the superconducting
cotunneling involving the transfer of the electron on dot 2 to dot
3 through the SC electrode R (involving the $(1,1)$ charge state) and then the transfer of the electron
on dot 1 to dot 2 through the SC lead L;  however, this would only renormalize $|{\cal T}'|$.
 Other (parasitic) SC cotunneling terms can be completely ignored through the prerequisite $U-\epsilon\gg \epsilon$  (and $(E_{20},E_{02})\gg E_{00}$).
Below, we build explicitly the effective ${\cal H}^{eff}$ in terms of the phase $\varphi =\phi_L({\bf r}_L) - \phi_R({\bf r}_R) -\frac{\pi}{\Phi_0}\int_{{\bf r}_R}^{{\bf r}_L} (d{\bf l}_{13} + d{\bf l}_2)\cdot{\bf
A}$ (notice that the integration from ${\bf r}_R$ to ${\bf r}_L$ runs via dot
2 or dots 1 and 3):
\begin{eqnarray}
\label{H} {\cal H}^{eff} &=& {\cal H}_0+{\cal H}_{{\cal
T}'}+J\left( {\bf S}_2 \cdot {\bf S} - \frac{1}{4}
\right)\left(\hat{P}_{10}+\hat{P}_{01}\right) \\ \nonumber &+& J
e^{i\varphi-i\phi({\bf r}_{1}-{\bf r}_3)}\left( {\bf S}_2
\cdot {\bf S} - \frac{1}{4} \right)\hat{Q}^{+} \hat{P}_{10} \\
\nonumber &+& J e^{-i\varphi+i\phi({\bf r}_{1}-{\bf r}_3)}\left(
{\bf S}_2 \cdot {\bf S} - \frac{1}{4} \right)\hat{Q}^{-}
\hat{P}_{01}.
 \end{eqnarray}
Remember that $\phi({\bf r}_{1}-{\bf r}_3)=-\frac{\pi}{\Phi_0}\int
_{{\bf r}_3}^{{\bf r}_1} d{\bf l}_{13}\cdot {\bf A}$ depicts the
Aharonov-Bohm phase accumulated to go from dot  3 to dot 1,
$\varphi-\phi({\bf r}_1-{\bf r}_3)$ does not depend on $\lambda$,
and by construction $J>0$. Here,  ${\bf S}=\sum_{\alpha\beta}
d^{\dagger}_{\alpha}\frac{{\bf \sigma}_{\alpha\beta}}{2}
d_{\beta}$ embodies the spin of the delocalized electron on dots 1
and 3. An entanglement occurs between the spin on dot 2 and the
spin ${\bf S}$ of the delocalized qubit. The process of Fig. (2b)
enables the singlets to resonate via ${\bf S}_2\cdot{\bf
S}|10\rangle \rightarrow {\bf S}_2\cdot{\bf
S}\hat{Q}^+\hat{P}_{10}|10\rangle ={\bf S}_2\cdot{\bf
S}|01\rangle$. Since ${\cal H}_{\cal T'}$ does not involve spin
degrees of freedom this does not affect the spin entanglement. On
the contrary, through a flip $\hat{P}_{10}\rightarrow
\hat{P}_{01}$ this also stimulates the resonance of the singlets.
The antiferromagnetic exchange coupling reads
 \begin{eqnarray}
J(\delta r)= \frac{2\Gamma^2}{\epsilon+E_{00}-E_{Q}}  \left(\frac{\sin(k_F \delta r)}{k_F \delta r}\right)^2 \exp(-2\delta r/\pi \xi),
\end{eqnarray}
where $\Gamma=\pi |{\cal T}|^2 {\cal N}(0)$ with ${\cal N}(0)$
being the normal state density of states per spin of the leads at
the Fermi energy ${\epsilon}_F>\Delta$ and ${\bf k}_F$ $(k_F=|{\bf
k}_F|)$ being the Fermi momentum. Note that the lowest-order
expansion (4) in ${\cal H}_{\cal T}$ is valid in the limit where
$\Gamma\ll \epsilon + E_{00}-E_{Q}$. Here, $\xi=v_F/(\pi \Delta)$
with $v_F$ the Fermi velocity represents the coherence length of
the SC leads. Remember that $\delta r$ denotes the typical
distance between two injected spin-entangled electrons in a given
SC lead in the case where one electron stems from dot 2 and the
other from, say, dot 1. Thus, to have $J(\delta r)$ non-zero,
$\delta r$ should not exceed the SC coherence length. For
conventional s-wave superconductors the coherence length $\xi$ is
on the order of
\begin{figure}[ht]
\begin{center}
\includegraphics[width=5.9cm,height=4.3cm]{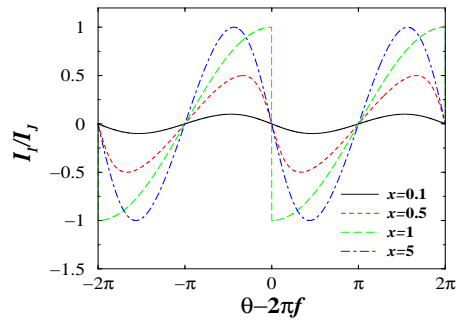}
\end{center}
\vskip -0.6cm
\caption{(Color online) $I_1/I_J$ versus $\theta-2\pi f$ for different values of $x=|{\cal T}'|/J$. When ${\cal T'}\sim 0$, $E^{eff}$ does not depend on $\phi_R-\phi_L$ resulting in $I_1=0$. When $|{\cal T'}|\gg J$, the sinus-like  profile is reminiscent of the double dot case\cite{Choi}. For $|{\cal T'}|=J$, when the interference  $|\langle \Psi_J|\Psi_Q\rangle|$ is fully destructive ($|\langle \Psi_J|\Psi_Q\rangle|\rightarrow 0$) and $\sin(\theta-2\pi f)$ changes  sign, an abrupt jump in $I_1$ occurs.}
\end{figure}
micrometers and thus this poses not severe experimental
restrictions. The suppression of $J(\delta r)$ is only polynomial
$\propto (1/k_F \delta r)^2$. Exploiting parameters for a
two-dimensional electron gas (InAs or GaAs) coupled to a
superconductor (Al or Nb) leads to $J\sim 0.05-0.5$K\cite{Choi}.
The ground-state energy (the lowest eigenenergy) for a singlet
state on the triple dot with $\langle {\bf S}_2 {\bf S} -
\frac{1}{4}\rangle =-1$ then can be evaluated in the charge
subspace $(|10\rangle,|01\rangle)$:
\begin{equation}
E^{eff} = E_Q-J-\epsilon -\left(J^2+|{\cal T}'|^2-2J|{\cal T}'|\cos(\varphi-\lambda)\right)^{1/2}.
\end{equation}
We emphasize that the mesoscopic RVB system might be detectable through the SQUID-ring setup in Fig. 1;  note that an excited triplet state on the triple dot leading to $\langle {\bf S}_2 {\bf S} - \frac{1}{4}\rangle =0$ does not contribute to the supercurrent.

{\it SQUID analysis}.--- The Josephson current through the junction reads $I_1=(2\pi c/\Phi_0) \partial E^{eff}/\partial(\phi_R-\phi_L)$.
When a Cooper pair accomplishes a roundtrip along the SQUID loop, we get the
phase sum rule $2\pi f = 2\pi \Phi/\Phi_0 = \theta+\varphi-\lambda$ where $\Phi$ is the flux threading
the SQUID loop, $\varphi-\lambda$ is the total (gauge-invariant) phase difference accross the triple dot, and $\theta\neq 0$ is the gauge-invariant phase difference accross the auxiliary junction $(J')$. Thus
\begin{equation}
I_1 = -I_J\frac{|{\cal T'}|\sin(\theta-2\pi f)}{\left(J^2+|{\cal
T}'|^2-2J|{\cal T}'|\cos(\theta-2\pi f)\right)^{1/2}},
\end{equation}
with the critical current $I_J=2eJ/\hbar\sim 5-50nA$. Note, in
Fig. 1, $I_S=I_1+(2eJ'/\hbar)\sin\theta$. Note also that $I_1$ in Eq.
(8) shows a resonance which is a distinct feature of this mesoscopic
RVB setup. Two limiting cases can be distinguished. When $|{\cal
T}'|\gg J$, the dots 1 and 3 form a unique effective grain. Here
$I_1=-I_J\sin(\theta-2\pi f)$ resembles the supercurrent through a
double dot in a vertical configuration\cite{Choi}; the extra minus
sign results from the interplay between ${\cal H}_{{\cal T}'}$ and
the Andreev process of Fig. 2b. When ${\cal T}'=0$, we infer that
$I_1=0$ (up to negligible contributions from paths involving only
dot 2\cite{Choi}). The dots 1 and 3 are still coupled through the
Andreev process of Fig. (2b) that is described by the phase
$\varphi-\phi({\bf r}_1-{\bf r}_3)$. Thus, projecting Eq. (5) on
the singlet state of the triple dot, the charge qubit is embodied
by the ground-state wave function $|\Psi_J\rangle =
\sqrt{P_{10}}|10\rangle+\sqrt{P_{01}}e^{i(\varphi-\phi({\bf
r}_1-{\bf r}_3))}|01\rangle$ with $P_{10}=P_{01}=1/2$ at
resonance. This gives $E^{eff}=\langle \Psi_J| H^{eff}
|\Psi_J\rangle = E_Q-\epsilon-2J$. The charge qubit seeks to react
by minimizing the energy independently of $\phi_R-\phi_L$.

Now, let us discuss the highly-resonating situation $|{\cal
T}'|=J$. Owing to $|\langle
\Psi_J|\Psi_Q\rangle|^2=[1-\cos(\theta-2\pi f)]/2$,
$\frac{I_1}{I_J}\approx -\sin(\theta-2\pi f)/(2|\langle
\Psi_J|\Psi_Q\rangle|).$ For half-integer values of $f$ if
$\theta$ is around zero, we recover a conventional sinus-like
behavior $I_1=-I_J|{\cal T'}|\sin(\theta-2\pi f)/(J+|{\cal
T'}|)\sim -I_J\sin(\theta-2\pi f)/2$. On the other hand, for
$(\theta-2\pi f)\approx 0$ or integer values of $f$, we observe
that $|\langle \Psi_J|\Psi_Q\rangle|\rightarrow 0$. Such a
destructive ``interference'' effect between ${\cal H}_{{\cal T}'}$
and the Andreev process of Fig. (2b) results in a marked jump in the
supercurrent (Fig. 3). The denominator in Eq. (8) becomes zero
whereas $\sin(\theta-2\pi f)$ changes sign. Jumps in $I_1$ only emerge in the highly-resonating regime
$|{\cal T}'|=J$ for the singlets and around $\theta-2\pi f=2\pi n$
($n$ is an integer) where $I_1=-I_J
sgn(\theta-2\pi f-2\pi n)$. For $\theta$ close to zero and
$\Phi=0$, the supercurrent yields $|I_1|=I_J$; this stands for a
distinct hallmark of this triple dot setup.

{\it Charge qubit off resonance.}--- For a finite energy splitting
$\delta E=E_{01}-E_{10}>0$, we must add the extra term $\delta
E\hat{Q}_z=\delta E(\hat{P}_{01}-\hat{P}_{10})/2$ in ${\cal
H}^{eff}$. When $\delta E\ll (|{\cal T}'|,J)$ we expect that this
does not affect much the preceding results. On the contrary, for
$\delta E\gg (|{\cal T}'|,J)$, to minimize energy the double-dot
charge qubit will satisfy $P_{10}=1$ and $P_{01}=0$. Here,
$|01\rangle$ becomes a virtual (excited) state and this will
obviously hinder the resonance of the singlets on the triple dot.
For instance, $P{\cal H}_{{\cal T}'}P=0$ and
\begin{eqnarray}
{\cal H}^{eff} &=& {\cal H}_0+\delta E\hat{Q}_z+J\left( {\bf S}_2
\cdot {\bf S} - \frac{1}{4} \right)\hat{P}_{10}
\\ \nonumber
&-& \frac{2|{\cal T}'|}{\delta E} J\cos(\varphi-\lambda)\left(
{\bf S}_2 \cdot {\bf S} - \frac{1}{4}
\right)\hat{Q}^-\hat{Q}^+\hat{P}_{10}.
\end{eqnarray}
Since $\hat{Q}^-\hat{Q}^{+}\hat{P}_{10}=\hat{P}_{10}$ we check
that the spin entanglement now always involves the spin of the dot
1. We had to go one step further in the expansion of
the effective Hamiltonian in Eq. (\ref{EffH}) by including a final
tunnel event from dot 3 to dot 1. There is another allowed process
where the electron on dot 1 first tunnels onto dot 3, hence the
electrons from dots 2 and 3 leave into the SC electrode R.
Finally, a Cooper pair leaves the SC lead L allowing to return to
the state $|10\rangle$. The singlet contribution to $I_1$ reads
$-I_J(2|{\cal T}'|/\delta E)\sin(\theta-2\pi f)$. By analogy to
the double-dot\cite{Choi}, the supercurrent profile is 
sinus-like.

{\it Conclusion}.--- We have theoretically introduced a triple dot that is a
mounting between a double-dot charge qubit and a spin qubit. By
placing the triple dot between two SC electrodes, a spin
entanglement can emerge between the spin qubit and the spin of the
delocalized electron on the double-dot charge qubit. When the
charge qubit is at resonance, the singlets propagate freely on the
triple dot owing to the equivalence between the charge states
$(1,0)$ and $(0,1)$ resulting in a mesoscopic RVB system and clear
predictions. When the resonance of the singlets is maximized, {\it i.e.}, for
$|{\cal T'}|=J$, we predict jumps in the supercurrent occurring for integer values of $f=\Phi/\Phi_0$ if
$\theta$ is around zero. To reach $|{\cal T}'|=J$, one could tune the tunnel coupling beween
dots 1 and 3 similar to Refs. \onlinecite{Petta,Jeong}. Since supercurrents can flow through a high-mobility AlGaAs/GaAs two-dimensional electron gas\cite{Marsh}, our proposal might be realized in the future.

{\it Acknowledgments}.--- We acknowledge M.-R.  Li and E. Sukhorukov for discussions.
This work is supported by CIAR, FQRNT, and NSERC (KLH), by the Swiss NSF, NCCR Nanoscience, EU Spintronics, and DARPA (DL), and by the University of Tokyo, JST, and
NTT (PR).
%

\end{document}